# Phase Sensitive Adiabatic States, or Does the Material Phase Recognize the Physical Reality?

I. G. Koprinkov

*Technical University of Sofia, Institute of Applied Physics, 1756 Sofia, Bulgaria*

**Abstract**

   Phase sensitive adiabatic states for a quantum system interacting with an electromagnetic field have been derived taking into account all material phase factors of the initial bare states. The adiabatic states so obtained show a traceable phase behavior, causally depending on the initial conditions and the relevant physical processes. Experimental appearance of the material phase effects has been discussed.

## 1. Introduction

   According to quantum mechanics, the physical state of a quantum system (QS) is most completely specified by the corresponding state vector. Although Schrödinger's equation determines the evolution of the entire state vector, physical meaning, within the standard interpretation of quantum mechanics, has only its modulus. Thus, any constant (in the configuration space) phase factor of unit modulus (while, otherwise, function of time) is considered, in principle, as unobservable [1]. In addition, due to the mathematical formalism, the physical state remains indeterminate up to an arbitrary true constant phase factor of unit modulus considered as irrelevant to the physical processes [2]. Formally, the constant phase factor can be incorporated in the definition of the state so that it is normally disregarded from the considerations. The relative constant phase of the state vectors is subject to limited attention only, *e.g.*, [3, 4], concerning the interference of wave functions and the quantum measurement process. In any case, however, the phase of the state vector is considered as *uncontrollable*, *unpredictable*, and *unobservable* quantity, which will be referred shortly to as $U^3$-*feature*.

   The total phase $\Phi_t$ of the state vector (material phase) has two contributions, different in their nature but equivalent in appearance in the state vector, one from the dynamical phase $\Phi_d$ and other from the geometric (Berry's) phase $\Phi_g$. They are defined as [5, 6]

$$\Phi_t = \arg\langle \Psi(0) | \Psi(t) \rangle \qquad \qquad 1(a)$$

$$\Phi_d = -\hbar^{-1} \int_0^\tau \langle \Psi(t) | \hat{H}(t) | \Psi(t) \rangle dt \qquad \qquad 1(b)$$

$$\Phi_g = i \oint_C \langle \Psi_n(s) | \nabla_s \Psi_n(s) \rangle ds \qquad , \qquad 1(c)$$

where $|\Psi_n(s)\rangle$ are basic states that evolve along a closed path $C$ in the parameter space, and $s$ is parameter(s) on which the Hamiltonian $\hat{H}$ depends.

As has been summarized [6], the Berry's phase has observable consequences in wide range of physical processes. If so, does the dynamical phase have also observable physical consequences? Some recent results, despite not considered in the sense discussed here, provoke us to state again this question. The phenomenological appearance of the dynamical part of the material phase (MP) could be distinguished in some experiments with material wave-packets (WPs) within atoms [7] and molecules [8-10], or with atomic de Broglie WPs [11] - the Feynman's *gedanken* experiment, using advanced optical technology. As the WPs are created by short optical pulses, in the time between pulses the WPs are ruled solely by the atomic/molecular Hamiltonian and the acquired phase is purely material. The fluorescence-detected interferograms substantially depend on the phase-locked optical frequency and, thus, the carrier WP (vibration) frequency [8, 9], which determines the acquired MP. Even the constant phase shift of the phase-locked laser pulses, creating the material WPs, leads to a dramatic change in the observed result [7-9]. In some cases [12], a constant, $\pi$-phase shift of the wavefunction is invoked to explain different physical behavior of the QSs - the reflected and transmitted atoms in the Bragg reflection from standing light field. All these inspire the understanding that, in principle, the MP may cause observable physical consequences at properly designed experiments.

From the other side, the influence of the optical phase on the atomic phenomena is well established both theoretically [13-16] and experimentally [17-20]. The rapid field phase fluctuations have been also considered in that sense [21-24]. While the optical phase is clearly separated from the MP discussed here, they both equivalently participate in the corresponding analytical expressions (see bellow). This will wash out the relevant dependence of the physical processes on the optical phase if the $U^3$-*feature* of the MP holds.

The purpose of the present work is to investigate the behavior of the dynamical part of MP when involve the QS in a real physical process, which, in principle, allows to trace the MP dynamic. For such a purpose we reconsider the basic problem of interaction of a QS with an electromagnetic field (EMF), paying special attention to the MP. We have found that the MP of adiabatic states (ASs) of the combined atom-field system shows traceable and physically sensible behavior, depending on the initial conditions and the relevant physical processes involved. These states were called phase sensitive adiabatic states (PSASs) (For the case of monochromatic field, see Ref. [25]). Some general problems around the experimental observation of the MP effects are also discussed.

**2. Phase sensitive adiabatic states**

All observable quantities we know at present are based on the respective matrix element, bilinear product of state vectors, which results in appearance of the phase difference of state vectors. Formally, the dynamical phase results also from bilinear product of state vectors, Eq. (1b). In a number of realistic cases, however, the phase (and amplitude) of one of the state vectors only carries the important physical information whereas the other state, subject to minor perturbations, practically stays unchanged while only its phase acquires "idle running" in time. This is the case, *e.g.*, when "upward" (excitation [10], ionization [7]) or "downward" (fluorescence) transition from the WP superposition state to another still empty eigenstate is used to probe the WP interference. This allows, formally, predetermining the phase so that the phase difference equals the absolute phase of one of the states - the state of the WP. That is why, the absolute value of the MP and the MP difference will be considered here as having same degree of importance. Knowledge of the absolute value of the MP refers the problem to the absolute internal clocking of the state of the QS. Since this problem seems intractable within the present experimental technique, it is more reasonable to talk about a relative phase, or a phase difference. Relative internal clocking of atomic and molecular states has been demonstrated by phase-locked [7-10] and phase-arbitrary [26] optical pulses. In this Letter we consider the constant (initial) and the time-dependent parts of the dynamical phase at an equivalent ground because of formal and phenomenological reasons: *i)* they have equivalent appearance in the initial and subsequent mathematical expressions, *ii)* they both behave as constants with respect to the Hilbert space, whose vectors are functions on configuration space, only, *iii)* the constant phase actually represents the time-dependent phase at the initial moment of creating the state, *iv)* changing the phase by a constant value has a dramatic effect on the interference phenomena with material WPs [7-9].

With the present analysis we aim at including as many as possible physical factors affecting the total phase of the state vector, while still remaining within the adiabatic conditions so as to keep the problem solvable analytically. Despite the mathematical complications in comparison with perfect adiabatic/monochromatic case, this allows making parallel between different phase contributions.

The ASs arise from the interaction of a QS with an EMF, see *e.g.*, [27]. They are eigenstates of the total Hamiltonian of the QS, $\hat{H}=\hat{H}_o+\hat{H}'$ and represent a more adequate basis for qualitative considerations of the field-matter interaction physics, than the states of the bare QS. The ASs (semiclassical dressed states) represent sum of products between the respective field and matter states, *i.e.*, field-matter entangled states. Thus, the ASs incorporate by a natural way all phase factors coming from the field and matter and represent very suitable tool in seeking of material phase effects. However, as the field and matter purely equivalently participate in the interaction and in the ASs, as well, it is more naturally to speak about unified field-matter phase effects.

Consider the basic model of two-level QS, (Fig.1), subject to near-resonant electrodipole interaction with a linearly polarized EMF

$$E = (1/2) E_o(t) \ [\exp(i\Phi) + \exp(-i\Phi)] \quad , \qquad (2)$$

where $\Phi = \omega\, t + \varphi\,(t)$ is the optical phase. We derive the PSASs for QS driven by weakly nonadiabatic EMF in presence of dumping. The field nonadiabaticity includes relatively smooth amplitude $E_o^{-1}\partial_t E_o$ and phase $\partial_t \varphi$ variations, which, together with their time-derivatives ($\partial_t^n \equiv \partial^n/\partial t^n$), are much smaller than the product of the respective powers of complex frequency detuning $(\Delta\omega - i\gamma_e/2)$ and Rabi-frequency $\Omega$, *i.e.*,

$$\left|\partial_t^n(\partial_t\varphi - i\Omega^{-1}\partial_t\Omega)\right| \quad << \quad \left|\Delta\omega - i\gamma_e/2\right|^{n+1-k}\left|\Omega\right|^k \quad , \tag{3}$$

where $n = 0, 1, 2,...$, and $k = 0, 1, 2,..., n+1$. The generalized adiabatic conditions (3) unify and extend the adiabatic condition $\left|(\Delta\omega - i\gamma_e)^{-1} E_o^{-1}\partial_t E_o\right| << 1$ [27], and the Born-Fock adiabatic condition $\left|\partial_t\Omega^{-1}\right| << 1$ [28]. It must be underlined that the adiabatic conditions (3) can be satisfied even at short pulse excitation ensuring enough frequency detuning from the exact resonance and high field intensity (Rabi-frequency).

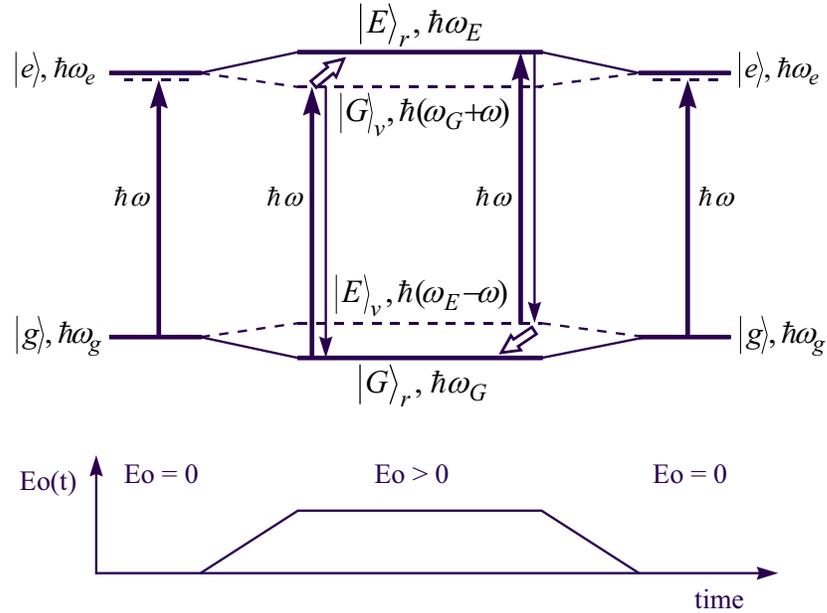

Fig.1: Bare ($E_o = 0$) and adiabatic ($E_o > 0$) states of the quantum system driven by an electromagnetic field. The bold arrows show the optical pumping, the empty arrow show the nonadiabatic processes.

When the EMF is switched-on, the QS will, in general, be in a superposition (coherent) state $\left|\Psi(\vec{r},t)\right\rangle$:

$$\left|\Psi(\vec{r},t)\right\rangle = g(t)\left|g(\vec{r})\right\rangle\exp[-i\Phi_g(t)] + e(t)\left|e(\vec{r})\right\rangle\exp[-i\Phi_e(t)] \quad , \tag{4}$$

where $g(t)$ and $e(t)$ are time-dependent amplitudes of the ground $\left|g\right\rangle$ and excited $\left|e\right\rangle$ bare states, $\hat{H}_o\left|g\right\rangle = \hbar\omega_g\left|g\right\rangle$, $\hat{H}_o\left|e\right\rangle = \hbar\omega_e\left|e\right\rangle$; $\Phi_g(t) = \varphi_g + \omega_g t$, $\Phi_e(t) = \varphi_e + \omega_e t$ are the total dynamical

5phases; $\omega_g$, $\omega_e$ are eigenfrequencies; and $\varphi_g$, $\varphi_e$ are the respective initial phases, which will be retained here due to above arguments.

To derive the PSASs, we solve the equations of motion for the amplitudes $g(t)$ and $e(t)$ of the state-vector (4) that result from the Schrödinger's equation $\hat{H}|\Psi(\vec{r},t)\rangle = i\hbar \partial_t |\Psi(\vec{r},t)\rangle$. Using rotating-wave approximation, and including a phenomenological dumping term, we have:

$$\begin{aligned}\partial_t g(t) &= (i/2)\Omega(t)\exp[-i\Delta\Phi(t)]\, e(t) \\ \partial_t e(t) &= -(1/2)\gamma_e\, e(t) + (i/2)\Omega(t)\exp[i\Delta\Phi(t)]\, g(t)\end{aligned} \quad , \tag{5}$$

where $\Omega(t) = \mu E_o(t)/\hbar$ is the resonance Rabi frequency, $\gamma_e = \gamma'_e - i\gamma''_e$ is the dumping rate of the excited state to other but not ground state $|g\rangle$ accounting for both, the shift ($\gamma''_e$) and the broadening ($\gamma'_e$) of the energy level due to the macroscopic environment, and $\Delta\Phi = \Phi_e - \Phi_g - \Phi$ is the phase difference. The association of ground state to the states to which the excited state may dump leads to mathematical complications only but does not change the main physical conclusions.

After eliminating $e(t)$ from Eqs.(5), one obtains

$$\partial_t^2 g + i\Delta\tilde{\omega}'(t)\partial_t g + (1/4)\Omega^2(t)\, g = 0 \quad , \tag{6}$$

where $\Delta\tilde{\omega}' = \Delta\omega - \partial_t\varphi - \gamma''_e/2 - i(\gamma'_e/2 - \Omega^{-1}\partial_t\Omega)$, and $\Delta\omega = \omega_e - \omega_g - \omega$ is the zero-field frequency detuning. Substituting $g(t) = f(t)\exp[-(i/2)\int_0^t \Delta\tilde{\omega}'dt']$, Eq.(6) can be expressed in the normal form,

$$\partial_t^2 f + (1/4)\tilde{\Omega}'^2 f = 0 \quad , \tag{7}$$

where $\tilde{\Omega}' = [\Delta\tilde{\omega}'^2 + \Omega^2 - i2\partial_t\Delta\tilde{\omega}']^{1/2}$ has meaning of instantaneous off-resonance Rabi frequency. In the general case, Eq.(7) does not have an exact analytical solution. Following [29], an approximate solution for the state vector $|\Psi(\vec{r},t)\rangle$ has been obtained [30] within the adiabatic conditions (3).

Even in the present rather complicated case, we can construct an ASs basis, neglecting (within the adiabatic condition (3)) the corresponding small terms from the solution for the state vector $|\Psi(\vec{r},t)\rangle$. Thus, the ground $|G\rangle$ and excited $|E\rangle$ ASs can be expressed as usual

$$\begin{aligned}|E\rangle &= COS(\theta/2)|E\rangle_r - SIN(\theta/2)|E\rangle_v \\ |G\rangle &= SIN(\theta/2)|G\rangle_v + COS(\theta/2)|G\rangle_r\end{aligned} \quad , \tag{8}$$

where the indexes "r" and "v" stand for their real and virtual components, Fig.1. The explicit form of the ASs components depends on the imposed initial conditions. At ground state initial conditions (initially, the QS is with certainty in the bare ground state $|g\rangle$) the real and virtual components of the ASs are:

$$
\begin{aligned}
|G\rangle_r &= |g\rangle \exp(-i\varphi_g) \exp(-i\int_0^t \omega_G dt') \\
|G\rangle_v &= |e\rangle \exp[-i(\varphi_g + \varphi)] \exp[-i\int_0^t (\omega_G + \omega)dt'] \\
|E\rangle_r &= |e\rangle \exp[-i(\varphi_g + \varphi)] \exp(-i\int_0^t \tilde{\omega}'_E dt') \\
|E\rangle_v &= |g\rangle \exp(-i\varphi_g) \exp[-i\int_0^t (\tilde{\omega}'_E - \omega)dt']
\end{aligned}
\qquad (9)
$$

where $\tilde{\omega}'_E = \omega_E - \partial_t\varphi - \gamma''_e/2 - i(\gamma'_e/2 - \Omega^{-1}\partial_t\Omega)$ is the effective frequency of the real excited state; $\omega_G = \omega_g + \Lambda_-$ and $\omega_E = \omega_e - \Lambda_-$ are the Stark-shifted frequencies of the real ground and excited states, respectively; $\Lambda_\pm = 1/2(\Delta\tilde{\omega}' \pm \tilde{\Omega}')$; $COS(\theta/2) = (\tilde{\Lambda}'_+/\tilde{\Omega}')^{1/2}$ and $SIN(\theta/2) = \text{sgn}(\Delta\omega)(-\tilde{\Lambda}'_-/\tilde{\Omega}')^{1/2}$ are intensity dependent "weight" factors (partial representation of the ASs components in the entire AS [31]); and $\tilde{\Lambda}'_\pm = \Lambda_\pm - (i/2)\tilde{\Omega}'^{-1}\partial_t\tilde{\Omega}'$. The relevant phase factors were separated in (9) in order to form the "energies" of the respective ASs' components. The solutions (9) generalize naturally the perfect adiabatic/monochromatic field solutions [25]. They can be reproduced from (9) eliminating all nonadiabatic field factors and dumping terms.

Eqs. (9) clearly show that the *material* phase and the classical *field* phase (the physical meaning of later is indisputable) participate *equivalently* in the formation of the total phase of the ASs of the combined atom-field system. Also, the ground state constant MP $\varphi_g$ appears in all AS components, whereas the excited state MP $\varphi_e$ totally disappears. At excited state initial conditions a similar solution takes place [30], but this time it is the excited state MP $\varphi_e$ appearing in all ASs components, whereas $\varphi_g$ totally disappearing [32]. Thus, the *material phase* behaves as a *traceable quantity* in the formation of the ASs. To the best of our knowledge, such a feature is formulated for the first time [25], and we call it *material phase tracking* (MPT). The MPT is the first statement concerning dynamical MP that differs from the standard $U^3$-feature. Since the ASs derive from the bare states by continuous evolution forced by the EMF, the bare states represent actually the zero-intensity limit of the ASs. That is why, we consider that the MPT is a general feature of the quantum states at all. However, there is no sense to talk about the MPT within unperturbed bare states since it manifests itself when the QS is involved in a real physical process. Based on Eqs. (9) we can go further showing that the phase of the PSASs behaves as an *additive dynamical quantity* that closely follows the process of its formation. Really, starting from $|G\rangle_r$ (ground state initial conditions), Fig.1, the phase of the virtual component $|G\rangle_v$ results from the phase of the real component $|G\rangle_r$, adding the optical phase. At the same time, the state $|G\rangle_v$ results physically from $|G\rangle_r$ by temporal reversible absorption of one field photon. Next, due to the nonadiabatic coupling between $|G\rangle$ and $|E\rangle$, the phase of $|G\rangle_v$ transfers to $|E\rangle_r$ acquiring nonadiabatic contributions from both, the field ($\partial_t \varphi$, $\Omega^{-1}\partial_t\Omega$) and the dumping factors ($\gamma_e''/2$, $\gamma_e'/2$). The variation of the field phase $\partial_t\varphi$ affects the instantaneous energy of the level, whereas the variation of the field amplitude $\Omega^{-1}\partial_t\Omega$ affects the amplitude of the state, populating $|E\rangle_r$ from $|G\rangle_v$. This is accompanied by irreversible absorption of some of the temporally absorbed photons. Finally, the phase of the virtual component $|E\rangle_v$ results from that of $|E\rangle_r$ subtracting the optical phase. Physically this corresponds to the temporal reversible emission of one photon from $|E\rangle_r$ to the field, thus forming the state $|E\rangle_v$. This process, stimulated by the photon flux, will be called *virtual stimulated emission*. Starting from $|E\rangle_r$, a similar phase behavior can be found for the PSASs derived at excited state initial conditions. Consequently, the phase of the PSASs is *causally* related with the relevant physical processes. This broadens our understanding for the MPT from the initial phase to the overall dynamical phase of the state vector. That is why we call the ASs (8), (9) - PSASs. The MPT is considered as the most primary and, thus, the most simple appearance of the material phase. More complicated appearance of the MP is expected when involve the internal degrees of freedom of QSs. They will lead to additional MP acquired by the QS. Such an appearance of the MP can be distinguished in the phase sensitive experiments with intraatomic and intramolecular WPs [7 - 10]. A physically sensible behavior of the material phase, such as the MPT, and, as we recognize for the first time, the experimentally observed manifestation of the MP [7-11], inspires the understanding that *the overall dynamical phase is causally involved in the dynamic of QS*. This, however, is not surprising if we remind the original form of the wave function, $\Psi = R\exp(iS/\hbar)$ [33], where $S$ is the Hamilton-Jacobi function (physical action). Thus, the phase ($\Phi = -S/\hbar$), which appears to be *a missed parameter* within the standard approach, is even more directly related to the dynamic of QS than the amplitude $R$, that has only an ensemble averaged probability meaning. All these do not mean to abandon the *epistemological* meaning of the entire wave function (probabilistic interpretation of its amplitude), neither to support the *ontological* extreme, considering $\Psi$ as an objectively real field [33]. We consider that, while the entire wave function $\Psi$ has not physical meaning, its elements, *i.e.*, its amplitude $R$ and phase $\Phi$ (or $S$), which are not independent but codetermine each other [33], have observable appearance in the physical processes.

Transitions between different ASs result from nonadiabatic factors (nonadiabaticity of the EMF, and/or dumping due to the zero-point vacuum fluctuations, and collisions) acting on the QS. The field nonadiabaticity is explicitly accounted for by $E_o^{-1}\partial_t E_o$ and $\partial_t\varphi$ terms. They do not lead to unpredictable and uncontrollable phase contributions during the nonadiabatic transition because, due to the classical nature of the EMF considered here, its amplitudes and phase, and thus $E_o^{-1}\partial_t E_o$ and $\partial_t \varphi$, are simultaneously well determined. However, a real source of arbitrary phase contributions may result from the zero-point vacuum fluctuations and collisions, responsible for dumping. It can be described by a definite value of $\gamma'_e$ and $\gamma''_e$ only in an ensemble averaged sense, while, for single QS, the dumping has statistical appearance. Their contribution, however, can be done negligible for a time scale much shorter then the reciprocal dumping rates. The quantum nature of light may also lead to such effect by, formally, the number-phase uncertainty relation. The MPT and the arbitrary phase contributions are simply different aspects of the MP effects of the QS.

**3. Experimental appearance of the material phase**

In the following, some principle aspects in the experimental observation of the MP effects will be discussed based on an experiment with material WPs (hereafter referred to as WP-experiment), similar to those within atoms [7] or molecules [8-10], but emphasizing on the MP. When discussing that problem we have to bear in mind that many physical quantities cannot be affected, observed, or measured directly due to the experimental or methodological limitations. That is why, if we find that changing (directly or indirectly) of given quantity affects at least one quantity, subject of direct or indirect observations, we may conclude that the quantity under consideration is *causally* related with the *physical reality*, and will be called *physical quantity*. In principle, such *causality* can be indisputably established and, to our opinion, this is the most important problem that should be solved about given quantity. To such a quantity can be ascribed *physical meaning*. The existence of causal relationship with the physical reality is base of more or less convincing (precise) experimental observation (measurement) of given quantity. The above speculations represent a more "relaxed" understanding of the *physical reality - physical quantity* relationship than the definition given in Ref. [34], according to which, "every element of the physical reality must have a counterpart in the physical theory". While, undoubtedly, this would be the best, it is difficult to predict when such a program will be realized. Strictly speaking, we do not know all elements of the physical reality. Sometimes it is not even possible to guess the element of the physical reality without definite knowledge about the corresponding physical quantity coming from the experimental and theoretical studies. One may happen that nearly all knowledge about given element of the physical reality comes from our knowledge about the corresponding physical quantity and they cannot be well separated in our mind. Namely, the meaningful physical behavior of given physical quantity reveals its relation with the physical reality. This necessarily requires that the theory must be open for encompassing of new physical quantities or elements of the physical reality. Here, we require the new quantity to be causally related with at least one of the already known physical quantities (but not to, sometimes, abstract element of the physical reality) so as to be considered as related with the physical reality. What element of the physical reality corresponds to given quantity, if not apparent, is subject of further studies. The above approach seems working in practice and will be used here. It must be underlined that both approaches, the present one and that one of Ref. [34], do not contradict each other while follow different philosophy.

Consider short optical pulse creating a WP $\Psi_1(\vec{r},t_1) = A_1(\vec{r},t_1)\exp(-i\varphi_1)$ on a real energy level/surface $V_1$ at moment $t_1$, Fig.2, where $A_1(\vec{r},t_1)$ and $\varphi_1$ are the initial amplitude and phase, respectively. The phase $\varphi_1$ appears as insubstantial constant phase only if we ignore the history of the QS before $t_1$. Actually, the initial phase $\varphi_1$ is predetermined from the earlier physical processes. In accordance with the general behavior of the MP, Eqs. 9, the phase $\varphi_1$ of the WP $\Psi_1$ on the real excited state $V_1$ results from the instantaneous MP of former real state (*e.g.*, ground state $V_o$, not shown in the figure), from which the WP was excited, adding the (locked [7-9]) field phase and the nonadiabatic phase contributions, due to transition from the temporally created virtual state to the real state $V_1$. Thus, the instantaneous overall phase at the moment when the respective real state is just populated appears as initial phase of that state. Once the pump pulse terminates, the WP propagation on $V_1$ is ruled solely by the molecular/atomic Hamiltonian and the phase $\Phi$ acquired while this takes place is purely material. The amplitude $A_1$ also changes because the WP is not eigenstate of $V_1$. Some time $\tau$ later (usually, the time after which the WP returns back to the area it was created), the WP becomes $\Psi_1(\vec{r},t_1+\tau) = A_1(\vec{r},t_1+\tau)\exp[-i(\varphi_1 + \Phi(\tau))]$. Then, at the moment $t_2 = t_1 + \tau$, a second WP $\Psi_2(\vec{r},t_2) = A_2(\vec{r},t_2)\exp(-i\varphi_2)$ is created on $V_1$. Both WPs interfere leading to superposition state $\Psi$

$$\Psi = A_1(t_2)\exp[-i(\varphi_1 + \Phi(\tau))] + A_2(t_2)\exp(-i\varphi_2) \tag{10}$$

The population density of the superposition state $\Psi$ on $V_1$ is given by a standard interference expression

$$\Psi^*\Psi = A_1^2 + A_2^2 + 2A_1A_2\cos(\varphi_1 - \varphi_2 - \Phi) \tag{11}$$

As Eq. (11) shows, the population density can be ruled experimentally either by the optical phase (that implicitly participate in the determination of the initial phases $\varphi_1$ and $\varphi_2$), or by the purely MP $\Phi$ (that depends on the vibration energy of the WP, which can be altered by the frequency of the pump pulses). Gain or quench of the population has been experimentally observed [7-10], changing the phase of the WPs. The effect of the purely MP $\Phi$ on the WPs population interferogram, and, thus, on the fluorescence signal, has been *experimentally proved* [8, 9], changing the locked frequency of the pump pulses. This fact is sufficient to believe that the MP has observable physically consequences. According to the above definition, the MP must be considered as a *physical quantity*.

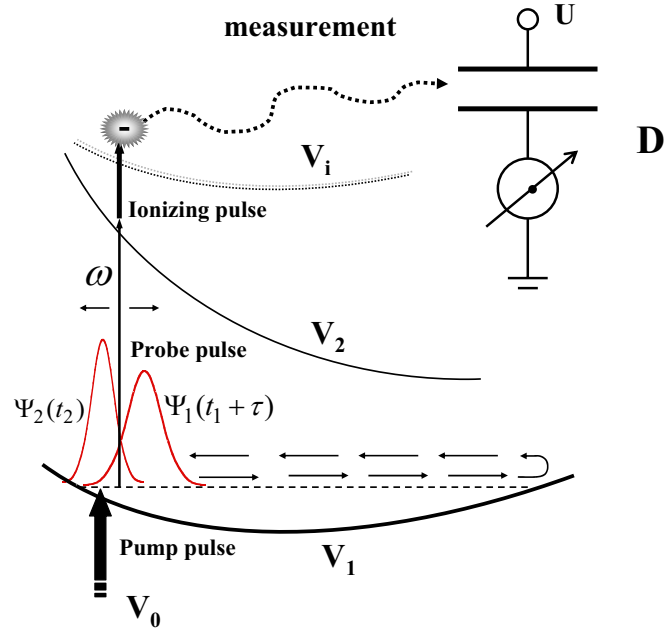

Fig.2: Experimental arrangement of material phase experiment. A sequence of two short pump pulses creates material wave-packets on the energy surface $V_1$. A weak probe (and ionizing) pulse "peaks up" the population without disturbing the coherent superposition on $V_1$. The created charge, proportional to the coherent superposition population, is detected by an incoherent detector D.

To achieve well expressed MP effects, the WPs (pump pulses) must be much shorter than the electron orbiting period in atoms [7] or nuclear vibration period in molecules [8-10]. Additional advantage of short pump pulses is that the nonadiabatic phase contribution of dumping can be done negligible, because the time interval between pulses, which normally equals the above periods, is much shorter than the lifetime of $V_1$. If we probe the population by ionization of the atoms/molecules by sequence of short probe-ionizing pulses, Fig.2, the arbitrary nonadiabatic phase contribution of dumping can be done negligible during the entire chain of measurement. Similar to the WP-experiment is the coherent control [13-20], where, however, the pump pulses, applied simultaneously in parallel to given transition, are much longer than the corresponding material time. The excitation is not localized inside the QS and the MP effects cannot be distinguished from the field phase effects. Nevertheless, the optical phase manipulation of the interference between transition amplitudes allows controlling the ionization yield or dissociation products in a real experiment [17-20].

The population can be probed by detecting the resonance fluorescence from $V_1$ [8, 9], from some higher excited state [10], or by ionization [7]. Usually, one detects signal proportional to the integral population in the overlapping area. The WP-experiment can be refined so as to allow observation of local features, approaching the population density. This can be achieved using a weak probe pulse that is close in resonance with some strongly repulsive molecule state $V_2$, Fig.2. This makes the detected

11signal highly sensitive from the probe pulse frequency. Actually, due to the wide bandwidth of the short pump pulses, the population density of the WPs occupies a number of rovibronic molecular levels. This requires a resonant probe pulse of "matched" frequency bandwidth. Changing the probe pulse frequency, the position where the resonant transition takes place moves left or right on the internuclear distance. This will allow to "pick up" population from different space positions, thus taking look inside the WP interference picture. Of course, physical limitations as, *e.g.*, the space distribution of the repulsive state vibration wave function, do not allow perfectly local probing of the population, or the population density. The maximum transition probability takes place around the "turning point" of $V_2$, where the vibration wave function have strong local maximum. The width of the later, which becomes narrower when the potential is steeper, restricts the spatial resolution of the population probe. While the local probe gives more detail picture of the WP interference, the total population change is sufficient for the qualitative observation of the MP effect.

An important feature in the WP-experiments is that, once the similarly prepared WPs superimpose, the population density from both WPs is no more distinguishable at given space position on the energy level. Consequently, whatever is the probe method, one detects the population created by both WPs simultaneously in the area of superposition.

**4. Discussion**

In the following, some particular questions concerning the PSASs and the WP-experiment will be discussed. In view of the various classifications of the quantum states, it will be useful to discuss first the PSASs from a terminological point of view. The PSASs were called *adiabatic* because an (generalized) adiabatic condition, Eq. (3), is required for their derivation. From the other side, the PSASs are eigenstates of the total field-matter Hamiltonian (the electromagnetic field is treated here classically), that is why, emphasizing on such a point, they can be considered as semiclassical *dressed* states. In contrast to other known adiabatic states, or semiclassical dressed states, *e.g.*, [27], the PSASs contain explicitly the time derivatives of the amplitude and the phase of the EMF, which represent field nonadiabatic factors. The other nonadiabatic factor is dumping, which also takes place in the PSASs. Similar to the present states (while not considering the initial MP as well as the applied here non-conventional interpretation) have been introduced and discussed in [35]. The later were called nonadiabatic semiclassical dressed states, emphasizing on the fact that they explicitly include the nonadiabatic factors. Finally, the PSASs, being dressed states, represent a sum of product (with definite "weight" factors) of the respective states of the interacting systems (in our case, the QS and the EMF), and, on such a ground, they can be considered as *entangled* states. Such structure of the dressed states is clearly expressed in the case of full quantum mechanical treatment of both, the field and the matter, while it can be also distinguished in the case of the classical treatment of the EMF. The PSASs, considered as entangled state, arise, in our case, between the ground state $V_0$ (not shown in Fig.2), from which the excitation starts, and excited state $V_1$ of the QS, from one side, and the (strong) *pump* field, from the other. Summarizing what has been said above, the most correct (and short) terminology that can be applied to the present states (while already called PSASs) is Phase Sensitive Nonadiabatic Dressed States. Although the terminology is a secondary question, such a clarification seems necessary in order to avoid some misunderstandings, as well as to underline the relationship of the PSASs with the nonadiabatic dressed states [35].

The WP-experiment should be clearly distinguished from some other experiments as, *e.g.*, the Feynman's "which-way" [11, 12] and the Einstein-Podolsky-Rosen (EPR) [34] *gedanken* experiments, in which the WP reduction takes place at given conditions according to the von Neumann's projection rule. In the "which-way" *gedanken* experiment, the resultant WP collapses onto some of the two interfering WPs if some of the pathways of the initial WPs become distinguishable alternatives as a result of given observation/measurement. In the WP experiment such two-way excitation exists within the $V_0$ - $V_1$ transition. Each of two pump pulses creates a WP on $V_1$, and both WPs coherently superimpose. However, as no measurement takes place on each of these two pump pulses, or pathways of creation of population from $V_0$ to $V_1$, no collapse of the resultant WP must be expected. Thus, the WPs' interference, including the MP information, is not destroyed. The probe and ionizing pulses simply "peak up" the total coherent superposition population from the interference of the WPs on $V_1$ (created by the pump pulses), and create a charge carrying information about that population, which finally affects the incoherent detector D. Consequently, no reason for reduction of the WP exists what concerns the existence of distinguishable alternatives/pathways.

In the EPR *gedanken* experiment, reduction of the WP and EPR type correlations occurs when perform a measurement on some of the subsystems of given combined system, being in an entangled state $\Psi$, so as to determine in which state this subsystem is. Then, the entangled state collapses onto the product of the respective eigenstates $\chi_\alpha$, $\pi_\alpha$ of both subsystems,

$$\Psi = \sum_i \chi_i \pi_i \rightarrow \chi_\alpha \pi_\alpha \qquad . \qquad (12)$$

In the arrangement of the WP-experiment, Fig.2, the (pump) field-matter system is in an entangled dressed state between $V_0$ - $V_1$ states of the QS and the EMF only until the preparation of the WPs under short pump pulses takes place. Once a definite state of this entangled state is created/populated (in our case, this is the excited state $V_1$), the entangled state between the QS and the EMF is destroyed and it collapses onto the excited state $V_1$ and the respective state of the EMF. The creation of such a definite state represents, in fact, the creation of the WP on $V_1$ by means of given pump pulse. After the WPs are prepared, they evolve on $V_1$ freely in time ruled only by the Hamiltonian of the QS itself and superimpose to create a coherent superposition of WPs. After the creation of the WPs, the pump pulses go away from the QS, leaving the QS is a well defined state - the excited state $V_1$, and the entangled dressed state between $V_0$ and $V_1$ states of the QS and the EMF cannot be created again. The probe and ionizing pulses are weak and assume to be far off $V_0$ - $V_1$ resonance, and are unable to create an entangled dressed state of significant amplitude. The probe and ionizing pulses simply "peak up" the superposition population on $V_1$ without creating a further collapse of state $V_1$, because it no more represents a superposition of some other states, different from $V_1$ (with its vibration and rotation substructure). What, actually, is superimposed on $V_1$, this is the population/WPs created on $V_1$. Thus, the measurement on the QS (which finally represents the created charge that hits the detector D) does not destroy the coherent superposition of the WPs/population. Such a measurement disturbs by no ways the interference picture inside the QS, which is important for the MP effects. Consequently, no WP reduction and no loss of (phase) information are expected to take place once the coherent superposition of the WPs is created by the pump pulses. This also means that, in the WP-experiment, no EPR type correlations [34] can be expected once the coherent superposition of WPs is created. Such a clarification is important because the entangled states (as the PSASs can be also considered) are usually associated with the reduction of the WP.

The present work may invoke associations with the "hidden variables" [33] and completeness [34] of the quantum mechanics. Although we recognize their importance, no direct relation with these problems in their standard meaning exists. The present work is performed within the quantum mechanics in the sense that the physical state is described by the state vector, which obeys the Schrödinger's equation. However, no restrictions from the standard, Copenhagen interpretation have been followed. This allows treating the problem of the MP based solely on the logical consequences from the obtained solutions as well as from the possibilities such consequences to be distinguished in a real experiment.

**5. Summary**

We have found for the first time that the material phase of a quantum system, involved in some real process, shows traceable physical behavior in the formation of the states of the quantum system. The above result, together with some phase-sensitive experimental results with material wave-packets [7-11], are evidences that the material phase is causally related with the dynamics of the quantum system. This states the question for physical reality beyond the standard interpretation of quantum mechanics.

**References**

1. J. von Neumann, *Mathematische Grundlagen der Quantenmechanik*, (Verlag von Julious Springer, Berlin, 1932), Chapter 1.
2. L. D. Landau and E. M. Lifshitz, *Quantum Mechanics, nonrelativistic theory* (in Russian), (Nauka, Moscow, 1989) p.20.
3. D. Bohm, *Quantum Theory*, (Prentice-Hall Inc., New York, 1952), Chapters 6 and 22.
4. J. Glimm and A. Jaffe, *Quantum Physics. A Functional Integral Point of View*, (Springer-Verlag, New York Inc., 1981), Chapter 1.
5. M. V. Berry, Proc. R. Soc. Lond., **A392** (1984) 45.
6. J. Anandan, Nature, **360** (1992) 307.
7. M. W. Noel and C. R. Stround, Jr, Phys. Rev. Lett., **75** (1995) 1252.
8. N. F. Scherer, A. J. Ruggiero, M. Du and G. R. Fleming, J. Chem. Phys., **93** (1990) 856
9. N. F. Scherer, et al., J. Chem. Phys., **95** (1991) 1487.
10. J. J. Gerdy, M. Dantus, R. M. Bowman and A. H. Zewail, Chem. Phys. Lett., **171** (1990) 1.
11. M. S. Chapman, et al., Phys. Rev. Lett., **75** (1995) 3783.
12. S. Dürr, T. Nonn and G. Rempe, Nature (London), **395** (1998) 33.

13. M. Shapiro, J. W. Hepbern, and P. Brumer, Chem. Phys. Lett., **149** (1988) 451
14. M. A. G. Martinez, P. R. Herczfeld, C. Samuels, L. M. Narducci, and C. H. Keitel, Phys. Rev. A **55** (1997) 4483
15. E. Paspalakis and P. L. Knight, Phys. Rev. Lett., **81** (1998) 293
16. Peng Zhou and S. Swain, Phys. Rev. Lett., **82** (1999) 2500.
17. C. Chen, Y. Yan and D. S. Elliot, Phys. Rev. Lett., **64** (1990) 507
18. S.-P. Lu, S. M. Park, Y. Yie and R. J. Gordon, J. Chem. Phys, **96** (1992) 6613
19. Qilin Wu, Daniel J. Gauthier and T. W. Mossberg, Phys. Rev. A **49** (1994) R1519
20. J. C. Camparo and P. Lambropoulos, Phys. Rev. A **55** (1997) 552.
21. G. S. Agarwal, Phys. Rev., **A18** (1978) 1490
22. B. J. Dalton and P. L. Knight, Opt. Commun., **42** (1982) 411
23. S. E. Harris, Phys. Rev. Lett., **70** (1993) 552
24. M. Fleischhauer, Phys. Rev. Lett., **72** (1994) 989.
25. For the case of monochromatic field see, I. G. Koprinkov, *Proceedings of the Int. Conference on Lasers'97, New Orleans, 1997*, ed. J. J. Carroll and T. A. Goldman (STS Press, 1998), pp. 291-297.
26. R. H. Bowman, M. Dantus and A. H. Zewail, Chem. Phys. Lett., **161** (1989) 297.
27. D. C. Hanna, M. A. Yuratich, D. Cotter, *Nonlinear Optics of Free Atoms and Molecules*, (Springer-Verlag, 1979).
28. N. B. Delone, and V. P. Krainov, *Atom in Strong Light Field* (in Russian), (Energoatomizdat, M. 1984), p. 64.
29. V. A. Kovarsky, N. F. Perelman, I. Sh. Averbukh, *Multiquantum processes* (in Russian), (Energoatomizdat, M. 1985), p.146.
30. I. G. Koprinkov, to be published.
31. Usually, the ASs are derived by suitable unitary transformation with transformation matrix elements - $\cos(\theta/2)$ and $\sin(\theta/2)$. Here, $COS(\theta/2)$ and $SIN(\theta/2)$ are complex quantities designate so as to underline their correspondence to $\cos(\theta/2)$ and $\sin(\theta/2)$, and, as they satisfy the formal condition $COS^2(\theta/2) + SIN^2(\theta/2) = 1$. The PSASs (9) do not form orthonormal basis because the basic Eqs.(5) of our non-conservative system with dumping correspond to non-hermitian Hamiltonian.
32. We have to underline that this is not a trivial elimination of the initial phases $\varphi_g(\varphi_e)$ by the initial conditions, as it is true also for the other elements of the initial state vectors. The initial conditions are imposed to determine unknown coefficients after the solution is found, so that they do not "destroy" mathematically the superposition state (4).
33. D. Bohm, Phys. Rev., **85** (1952) 166.
34. A. Einstein, B. Podolsky, and N. Rosen, Phys. Rev. **47** (1935) 777.
35. I. G. Koprinkov, J. Phys. B: At. Mol. and Opt. Phys., **34** (2001) 3679.